\documentclass[twocolumn,showpacs,preprintnumbers,amsmath,amssymb,prl]{revtex4}

\usepackage{graphicx}
\usepackage{dcolumn}
\usepackage{epsf}

\begin{document}

\title{Comment on: ``Vortex-assisted photon count and their magnetic field dependence in single-photon superconducting detectors" by L.\,N.\,Bulaevskii, M.\,J.\,Graf and V.\,G.\,Kogan, Phys. Rev. B {\bf 85}, 014505 (2012)}

\author{A. Gurevich$^{1}$ and V.M. Vinokur$^{2}$}
\affiliation{$^{1}$Department of Physics, Old Dominion University, Norfolk, VA 23529 \\
$^{2}$Materials Science Division, Argonne National Laboratory,
Argonne, IL 60439}

\date{\today}
\begin{abstract}

\end{abstract}

\maketitle
Recently L.\,N.\, Bulaevskii, M.\,J.\, Graf and V.\,G.\, Kogan published two papers on the theoretical description of
experiments on NbN thin film-based photon detectors~\cite{b1,b2}.  The central part of both works constituted the
calculation of the voltage produced by thermally-activated hopping of
vortices across a thin film strip. The authors of Ref.~\cite{b1,b2} used the approach developed in Ref.\,[\onlinecite{gv}], which, in turn,
was based on the earlier work by Ambegaokar et al.~\cite{amb} describing the dynamics of a vortex in a film in terms of the 
Langevin equation $\eta\dot{x}+U'(x)=\zeta(t)$. Here $x$ is the position of the vortex across the strip $(0<x<w)$,
$\zeta(t)$ is the thermal noise source, and $\eta$ is the viscous drag coefficient. The energy
$U(x)$ of a vortex in a film of the width $w<\Lambda=\lambda^2/d$ and the thickness $d$ in the presence of the dc
current $I$ and the magnetic field $H$ perpendicular to the film is given by~\cite{gv}
    \begin{equation}
    U(x)=\epsilon\ln\left[\frac{w}{\pi\xi_1}\sin\frac{\pi
    x}{w}\right]+\frac{\phi_0Ix}{cw}+\frac{\phi_0H}{8\pi\Lambda}x(w-x)\,,
    \label{u}
    \end{equation}
where $\epsilon=\phi_0^2/16\pi^2\Lambda$, $\xi_1\approx 0.34\xi$, $\xi$ is the coherence length,
$\lambda$ is the London penetration depth, $\phi_0$ is the flux quantum, and $c$ is the speed of light.
The voltage $V(T,H,I)$ caused by uncorrelated jumps of vortices is calculated using the Fokker-Planck
equation for the probability density $f(x,t)$ \cite{amb,fpe}:
\begin{equation}
\dot{f}=\partial_ x[U'(x)f+T f']
\label{fpe}
\end{equation}
The only difference between Eqs. (\ref{u}) and (\ref{fpe}) used in Refs.~\cite{b1,b2} and Ref.~\cite{gv} is that
the authors of Ref.~\cite{b1,b2} took $\xi_1=\xi/2$ in Eq.\,(\ref{u}), while in Ref.~\cite{gv} this quantity was
$\xi_1=0.34\xi$. The origin and the consequences of this discrepancy is discussed below.

The results of Refs.~\cite{b1,b2} essentially reproduce those of Ref.~\cite{gv}, however the
vortex hopping rate $R_v$ derived in Ref.~\cite{gv} differs by the factor $F=2^\nu(\nu-1)w/\xi$ from that of Ref.~\cite{b1,b2}.
The authors of Refs.~\cite{b1,b2} asserted that, for $\nu=\epsilon/T = 110$ used in Ref.~\cite{b1} to fit the experimental data, 
$R_v$ of Ref. \cite{gv} was overestimated by the factor $F=2^\nu(\nu-1)w/\xi\simeq 3.5\times 10^{36}$ because:

1. The factor $2^\nu$ in $F$ comes from the factor 2 under the logarithm in Eq.\,(\ref{u}) which, according to Ref.~\cite{b1,b2},
was missed in Ref.~\cite{gv}.

2. The factor $\nu -1$ in $F$ results from the use of the periodic boundary conditions for Eq.\,(\ref{fpe}) in Ref.~\cite{gv} as opposed to a 
`realistic' boundary condition of Ref.~\cite{b1}.

3. The factor $w/\xi$ in $F$ results from the vortex interaction which, according to Ref.~\cite{b1,b2},
 leads to the statistical weight of a vortex $P\sim L/w$, as opposed to $P\sim L/\xi$ used in~\cite{gv}, where $L$ is the length of the strip.

In this Comment we show that these statements are incorrect because they result from model artifacts of Refs.~\cite{b1,b2}.
Below we address the issues taken into account in Ref. \cite{gv} but neglected in Refs.~\cite{b1,b2} and discuss their importance 
for a more consistent theory of thermally-activated hoping of vortices in thin films and the interpretation of experimental data.

\subsection{1. Core contribution}

The authors of Ref. \cite{b1} apparently overlooked that $\xi_1=0.34\xi$ in Eq. (\ref{u}) of Ref. \cite{gv}
absorbs both the factor 2 under the logarithm \cite{b1,b2,stejic,vgk} and the
{\it vortex core energy} disregarded in Ref. \cite{b1,b2}. Here  Eq.\,(\ref{u}) results from $U(x)=\epsilon[\ln[(2w/\pi\xi)\sin(\pi x/w)]+\beta]$~\cite{stejic},
where $\beta = 0.38$ accounts for the vortex core energy, so that $\xi_1=0.34\xi = e^{-\beta}\xi/2$ in Eq.\,(\ref{u}),
unlike $\xi_1=\xi/2$ used in~\cite{b1,b2}.
The value $\beta=0.38$ was extracted by comparing the lower critical field $H_{c1}=(\phi_0/4\pi\lambda^2)[\ln(\lambda/\xi)+0.497]$
calculated from the Ginzburg-Landau (GL) theory~\cite{hu} with the London result $H_{c1}=(\phi_0/4\pi\lambda^2)[K_0(\xi/\lambda)+\beta]$,
where $\beta$ is the core contribution (associated with the spatial variation of the order parameter) not accounted for by the London cutoff,
and $K_0(\xi/\lambda)\approx \ln(2\lambda/\xi)-0.577$ for $\lambda \gg\xi$.
Matching these formulas for $H_{c1}$ yields $\beta = 0.497+0.577 - \ln 2=0.38$.
In the London limit, the core line energy is independent of the sample geometry so $\beta\approx 0.38$ is the same both in bulk samples and films with $w\gg \xi$, except for vortices spaced by $x\sim\xi$ from the surface.
Taking $\beta$ into account significantly decreases $R_v(\beta)\simeq \tilde{R}_ve^{-\beta\epsilon/T}$ as compared to $\tilde{R}_v$ calculated without the 
core contribution, while the variation of the core energy at the film edge affects the pre-exponential factor in $R_v$, as will be discussed below.

One can see that the core contribution $\beta=0.38$ is no less essential than the London numerical correction $\ln(2/\pi)=-0.45$ in $U(x)$, so 
taking $\beta$ into account is important when comparing the model \cite{b1,b2} with experiment.
Indeed, neglecting the vortex core energy in Ref.~\cite{b1} overestimates $R_v$ by $\sim\exp(\beta\epsilon/T)\sim  10^{18}$ for $\epsilon/T = 110$.
The importance of the vortex core contribution in the vortex-related dynamic phenomena has been extensively discussed in the
literature (see, for example, the recent work~\cite{terri} on the  effect of the vortex core energy on the Berezinskii-Kosterlitz-Thouless 
transition).

\subsection{2. Boundary condition}

Here we show that the extra factor $\sim \nu^{-1}$ in $R_v$ of Ref.~\cite{b1} does not come from the different
boundary conditions used in Refs.~\cite{b1} and~\cite{gv}, but rather from artifacts of the model of Ref.~\cite{b1}.
The vortex crossing rate was obtained in Ref.~\cite{b1} from the standard formula for a particle hopping between two potential wells~\cite{fpe}:
\begin{equation}
R_v^{-1}=D\int_0^{x_1}e^{-U(x)/T}dx\int_{x_0}^w e^{U(x)/T}dx,
\label{stoch}
\end{equation}
where $D=T/\eta$, $x_0\sim\xi$, and $x_1$ is a length smaller than
$x_m$ at which $U(x)$ is maximum.

The authors of Ref.~\cite{b1} assumed a model form of $U_{BGK}(x)$ in Eq.\,(\ref{stoch}): $U_{BGK}=U(x)$ where $U(x)$ is given by Eq.\,(\ref{u})
with $\xi_1=\xi/2$ for $x>x_0\sim\xi$, $U_{BGK}(x)=0$ for $0<x<x_0$, and $U_{BGK}(x)=\infty$ at $x=0$.
The infinite repulsive barrier at the film edge was introduced artificially to trap vortices in the film by imposing the boundary condition
of zero probability current $S={\dot x}f$ at $x=0$ for Eq.\,(\ref{fpe}).
Vortex hopping in this model occurs as a `pre vortex'~\cite{b1} is somehow placed in the film past this barrier,
but it is unclear how this model can describe penetration of vortices in the film.

The postulated form of $U_{BGK}(x)=0$ at $0<x\lesssim\xi$ significantly overestimates the first integral $Z=\int_0^{x_1} \exp[-U(x)/T]dx$ in Eq.\,(\ref{stoch}).
To see how it happens, it is instructive to juxtapose $U_{BGK}(x)$ with $U(x)$ obtained by numerical simulations of vortices using the GL equations,
which take into account the vortex core energy and its change near the edge. These calculations have shown that the energy of a vortex,
$U(x)= (b+ ax/\xi)\epsilon$, increases linearly with the distance $x$ of the core phase singularity from the film edge up to $x \sim \xi$ \cite{fertig,palacios}.
This gives rise to a constant force $a\epsilon/\xi$ caused by a ``string" of the suppressed order parameter between the core and the surface, where $a\sim 0.1-0.3$ and the constant $b\sim 0.05-0.1$ accounts for the fact that $U(x)>0$ even at $x \rightarrow 0$ due to local superconductivity suppression around a vortex core as it emerges from the film edge \cite{fertig}. These features are essential for the evaluation of $R_v$ if $U(x)>T$ at $x<\xi$.

Substituting $U(x)=(b+ax/\xi)\epsilon$ in $Z=\int_0^{x_1} \exp[-U(x)/T]dx$ yields $Z=\xi e^{-b\nu}/a\nu$ for $e^{-a\nu}\ll 1$. As follows from Eqs. 
(\ref{u}) and (\ref{stoch}), the factor $e^{-b\nu}$ can be combined with $e^{\beta\nu}$ from the second integral in Eq. (\ref{stoch}), so that 
the effect of the vortex core on the hopping rate  $R_v\simeq \tilde{R}_v\exp[(b-\beta)\nu]$ is determined by the difference of core energies in the bulk and at the film edge. Here both $a$ and $b$ appear to be dependent of current \cite{fertig}, indicating that the London notion of the rigid vortex core becomes hardly adequate at $x\sim \xi$.

The calculation of $Z\sim \int_{\xi_1}^\infty (\xi_1/x)^\nu dx \simeq \xi_1/\nu$ in Ref.~\cite{gv} is 
qualitatively consistent with the GL result. Here the cutoff $\sim \xi_1$ where the London theory becomes invalid was used,
and the upper limit can be extended to infinity if $e^{-a\nu} \ll 1$ and $I\ll I_d$,
where $I_d=c\phi_0/16\pi^2\Lambda\xi$ is of the order of the depairing current.
By contrast, the potential, $U_{BGK}(x)=0$ at $0<x<x_0$, yields $Z=x_0\sim \xi$ \cite{b1},
which overestimates $Z$ by the factor $\sim\nu\gg 1$ as compared to both the GL results and Ref.\,\cite{gv}.
Treating a vortex like a particle in the London model combined with the Fokker-Plank equation does bring uncertain factors
$\sim 1$ in $Z$ coming from the edge effects discussed above. Yet the simplified model of Ref.~\cite{b1} does not capture the 
qualitative behavior of $Z\sim \xi/\nu$, which follows from the more consistent GL theory and the approach of Ref. \cite{gv} (also adopted in Ref.~\cite{b2}).
A more realistic model of the vortex core penetration would require solving the time-dependent GL equations~\cite{tdgl}.

The above consideration shows that the claim of Ref.~\cite{b1} that the extra factor $\sim \nu$ in $R_v^{-1}$ comes from
the `realistic' boundary conditions as opposed to the periodic $U(x)$ of Ref.~\cite{gv} is misleading.
In fact, the solution of Eq.\,(\ref{fpe}) adopted in Ref.~\cite{gv} is only defined inside the film $0<x<w$ and does not require any unphysical barriers at the film edges. Here the use of periodic $U(x)$ in Eq.\,(\ref{fpe}) is a standard method of satisfying the boundary conditions of a fixed probability flux $S$ of vortices entering and exiting the film, which is equivalent to the method of images for solving the Laplace or diffusion equations. For example, Eq.\,(\ref{u}) can be obtained by either finding a proper analytical function or summing up potentials of an infinite chain of vortex-antivortex images outside the film.
Moreover, if only the forward jumps of vortices are taken into account in the limit of $e^{-\nu}\ll 1$ ~\cite{b1}, Eq.\,(\ref{stoch}) reduces to Eq. (7) of Ref.~\cite{gv}. This is not surprising because the exponentially small probability current $S$ is mostly determined here by narrow vicinities of neighboring minimum and maximum of $U(x)$, so the boundary condition of fixed $S$ \cite{gv} appears to be very close to the boundary condition $S=0$ of Ref.~\cite{b1}.

\subsection{3. Correlation effects}

Finally we comment on the statement of Ref.~\cite{b2} that the statistical weight of a single vortex penetrating through the film edge should be
$P\sim L/w$, instead of $P\sim L/\xi$ used in \cite{gv}. It is noteworthy that the models of Refs.~\cite{b1,b2,gv} only hold in the
limit of exponentially low density of vortices, thus $P$ should coincide with its value in the thermodynamic limit.
The assumption of $P\sim L/w$ is therefore inconsistent with the thermodynamics of vortices in thin films~\cite{amb,minhag} used to
obtain $P\sim L/\xi$ in Ref.~\cite{gv}. Here $P\sim  L/\xi$ is the 1D analog of the statistical weight $P = C(L/\xi)^2$ of a single vortex in the
film of area $L^2$ where $C\sim 1$ depends on the distribution of the order parameter in the vortex core~\cite{minhag}.

The assumption $P\sim L/w$ resulted from the interaction radius $\sim w$ of two vortices in the middle of the strip.
However, uncorrelated hopping of vortices described by Eq.\,(\ref{fpe}) imply that they enter the film at random times and are separated by distances
larger than $w$ at any given moment. Taking vortex correlations into account requires solving coupled equations for
the higher order correlation functions which cannot be described by Eq. (\ref{fpe}).
The authors of Ref. \cite{b2} selected rare events when two vortices enter the film nearly simultaneously and ascribed their statistical
weight $P\sim L/w$ to all vortex jumps. However, repulsion of vortices suppresses their simultaneous entering the film,
forcing them to go one by one so that a vortex can enter at any of $L/\xi$ edge sites after the preceding vortex in the area
$\sim w$ has already crossed the film. Such uncorrelated jumps \cite{gv} have a much higher
probability than the correlated jumps assumed in Refs. \cite{b1,b2}. In addition, the interaction radius of vortices at the film edge $(x\sim\xi)$ is 
much smaller than $w$ because currents of two vortices spaced by the distance $s$ along the edge are nearly extinguished by
their antivortex images, resulting in the dipole interaction $U(\xi,s)\sim \epsilon (\xi/s)^2$ which does not extend well beyond $s>\xi$.

In conclusion, we show the importance of the vortex core energy, the realistic behavior of $U(x)$ at the
film edge, and the physical boundary conditions to Eq. (\ref{fpe}) for the calculation of thermally-activated
hopping of vortices across narrow films. Disregarding these issues in Refs.~\cite{b1,b2} has produced large numerical
errors and a significant discrepancy between the results of  Refs.~\cite{b1,b2} and \cite{gv}.
This can also be essential for the interpretation of experimental data on thin film photon detectors.

This work was supported by the US Department of Energy Office of Science under the Contract
No.\,DE-AC02-06CH11357. AG was also supported by the Argonne National Laboratory 
through the subcontract NSC99-2911-I-216-001.



\begin{references}

\bibitem{b1}
L.\,N.\,Bulaevskii, M.\,J.\,Graf, and V.\,G.\,Kogan, \prb {\bf 85}, 014505 (2012).

\bibitem{b2}
L.\,N.\,Bulaevskii, M.\,J.\,Graf, C.\,D.\,Batista, and V.\,G.\,Kogan, \prb {\bf 83}, 144526 (2011).

\bibitem{gv}
A.\,Gurevich and V.\,M.\,Vinokur, \prl{\bf 100}, 227007 (2008).

\bibitem{amb}
V.\,Ambegaokar, B.\,I.\,Halperin, D.\,R.\,Nelson, and E.\,D.\,Siggia, \prb {\bf 21}, 1806 (1980).

\bibitem{fpe}
H.\,Risken, {\it The Fokker-Plank Equation}, Springer-Verlag, Berlin, Heidelberg, New York, Tokyo, 1984.
Eq. (\ref{stoch}) corresponds to Eq.\,(5.109) with $p=1$.


\bibitem{stejic}
G.\,Stejic, A.\,Gurevich, E.\,Kadyrov, D.\,Christen, R.\,Joynt, and D.\,C.\,Larbalestier, \prb{\bf 49}, 1274 (1994).

\bibitem{vgk}
V.\,G.\,Kogan \prb {\bf 49}, 15874 (1994); {\it ibid.} {\bf 75}, 064514 (2007).

\bibitem{hu}
C.-R.\,Hu, \prb {\bf 6}, 1756 (1972).

\bibitem{terri}
L.\,Benfatto, C.\,Castellani, and T.\,Giamarchi. ArXiv: 1201.2307v1


\bibitem{fertig}
D.\,J.\,Priour and H.\,A.\,Fertig, \prb {\bf 67}, 054504 (2003).

\bibitem{palacios}
P.\,S{\'a}nchez-Lotero and J.\,J.\,Palacios, \prb \textbf{75}, 214505 (2007)

\bibitem{tdgl}
A.\,N.\,Zotova and D.\,Y.\,Vodolazov, \prb {\bf 85}, 024509 (2012).


\bibitem{minhag}
P.\,Minnhagen, Rev. Mod. Phys. {\bf 59}, 1001 (1987).


\end{references}
\end{document}